\documentclass[fleqn,twoside]{article}
\usepackage{espcrc2}

\usepackage{graphicx}
\usepackage[figuresright]{rotating}
\newcommand{\e}{\epsilon}

\title{
\vspace{-1.5cm}
Numerical Evaluation of One-Loop Diagrams Near Exceptional Momentum
Configurations
\thanks{Talk given at the 7th DESY Workshop on Elementary Particle Theory, Loops and Legs 
in Quantum Field Theory, Zinnowitz 2004.}
\\
\vspace{-4cm}
\begin{flushright}\normalsize
FERMILAB-Conf-04/103-T\\
DCPT/04/72\\
IPPP/04/36\\
\vspace{2.5cm}
\end{flushright}
}
\author{W. Giele\address[fnal]{Fermilab, Batavia IL 60150,USA},
        E.W.N. Glover\address{Department of Physics, University of Durham,
        Durham DH1 3LE, England} and  G. Zanderighi\addressmark[fnal]}

\begin{document}

\begin{abstract}

One problem which plagues the numerical evaluation of one-loop Feynman
diagrams using recursive integration by part relations
is a numerical instability near exceptional momentum
configurations. In this contribution we will discuss a generic
solution to this problem. As an example we consider the case of forward
light-by-light scattering.
\vspace{1pc}
\end{abstract}

\maketitle

\section{Introduction}

By using recursion relations based on the integration by part techniques~\cite{CheTka} a general algorithm can be constructed to numerically evaluate
the finite part of one loop tensor $N$-point integrals~\cite{GiGlo}. The
numerical implementation is straightforward. However, for specific
configurations of the external momenta, the recursion relations build up a
numerical instability.  Phenomenologically, these exceptional kinematics are
characterized by such configurations as mass thresholds, forward scattering
and planar event where a
parameter tends to zero (for example the scattering angle). In this
contribution, we will discuss how to modify the recursion relations to produce
expansion formulae in the small parameter. 
This is an extension of the techniques developed in ref. \cite{CaGloMi}
(see also ref. \cite{FePaParUc})
and allows the evaluation of
tensor loop integrals at/near exceptional momenta configurations. 
As an example,
we work out in some detail forward light by light scattering through a massless
fermion loop: $\gamma\gamma\rightarrow \gamma\gamma$. The generalization to
massive particles with mass thresholds and processes with larger multiplicities
is in principle straightforward.  

\section{Exceptional Recursion Relations}

The dimensionality of space-time fixes the number of momentum vectors needed
to form a basis set in Minkowski space. 
As a consequence the integration by part
recursion relations fall into three categories depending on the number of external
legs, $N$, in the loop graph. These categories are: 
$N\leq 5$, $N=6$ and $N\geq 7$.

At the center of the recursion relations is the so-called kinematic matrix
\begin{equation}
S_{ij}=(q_i-q_j)^2-m_i^2-m_j^2\ ,
\end{equation}
where $q_i$ is the momentum flow through and $m_i$ the mass of propagator $i$.
For the three aforementioned categories of recursion relations, the kinematic
matrix has the following properties:
\begin{itemize}
\item[$N\leq 5$:]
Both the inverse of the kinematic matrix and the Gram determinant exist for
non-exceptional momentum configurations. The Gram determinant is proportional
to
\begin{equation}
B=\sum_i b_i=\sum_{ij} S^{-1}_{ij}\ .
\end{equation}
For momentum configurations close to exceptional kinematics the $B$
parameter tends to zero. Both the recursion relations and the basis set of loop
integrals
become numerically unstable in this limit. We need to setup a recursion relation
which generates an expansion in $B$.
\item[$N=6$:]
In this case, the inverse of the kinematic matrix still exists for non-exceptional
momentum configurations. However, $B=\sum S^{-1}_{ij}=0$ for all momentum
configurations. This leads to a different
set of recursion relations. Momentum configurations close to exceptional
kinematics are now characterized by small eigenvalues of the kinematic matrix.
In other words, the numerical determination of its inverse becomes unstable.
We need to setup a recursion relation which generates an expansion in the
small eigenvalue(s) of the kinematic matrix.
\item[$N\geq 7$:]
Here the kinematic matrix is singular for all momentum configurations. Note
that we do not need any special recursion relations since the standard recursion
relations are numerically stable.
\end{itemize} 
 
As we look at changing kinematic behaviour as the multiplicity
increases, we also see how to construct the recursion relations near exceptional
kinematics. Specifically, when we are exactly at an exceptional momentum configuration
for $N\leq 5$, i.e. $B=0$, we can use the $N=6$ recursion relations which 
were constructed on the premise that $B=0$. Similarly, for $N=6$, at the exceptional
momentum configurations the $N\geq 7$ recursion relations can be used.

However, we are interested in the phase space regions {\em close} to the exceptional
momentum configurations. It is clear how to construct the appropriate 
recursion relations in these regions. We need to rewrite the $N\leq 5$ recursion
relation as the $N=6$ recursion relation plus terms proportional to $B$ (i.e.
the Gram determinant). As we will see in the next section, this leads to an
expansion formula in $B$, enabling us to evaluate the loop graph near the exceptional
configurations with arbitrary precision. Equivalently, we can rewrite the
$N=6$ recursion relation into the $N\geq 7$ recursion relations plus
terms proportional to the small eigenvalue(s). This leads to
an expansion in the small eigenvalue(s).

\section{Forward $\gamma\gamma\rightarrow\gamma\gamma$ Scattering}

In this section we work out the simple example of light-by-light
scattering through a massless fermion loop. The analytic answers
for the helicity amplitudes are well known. Some helicity
amplitudes, such as 
$\gamma^+\gamma^+\rightarrow\gamma^-\gamma^-$ 
are constants, independent of the underlying kinematics.
Other helicity configurations give simple behaviour which is
logarithmically divergent as the scattering angle goes to zero.
Applying the recursion algorithm of Ref.~\cite{GiGlo} 
to this process leads
to a numerical evaluation of the amplitudes. However, when
the scattering angle becomes small the algorithm becomes numerically
unstable (see Figs.~4 and 5). 
The reason is twofold. First of all some of the 
recursion relations depend on the inverse of the $B$ parameter
\begin{equation}
B=\frac{2}{s}+\frac{2}{t}=-\frac{2u}{st}
=-\frac{2}{s}\left(\frac{1-\cos\theta}{1+\cos\theta}\right)
\end{equation}
which for small scattering angle $\theta$ tends to zero. Secondly,
one of the basis integrals, i.e. the end-point of the recursion relation,
becomes unstable. More precisely, the six-dimensional box integral has
a numerically unstable form (see Figs.~2 and 3):
\begin{eqnarray} 
\lefteqn{\lim_{sB\rightarrow 0}I(6;1,1,1,1)}\nonumber\\&=&
\!\!\!\!\lim_{sB\rightarrow 0}\frac{-1}{tsB}
\left[\log^2\!\left(\!1\!-\!\frac{sB}{2}\!\right)
\!-\!2\pi i\log\!\left(\!1\!-\!\frac{sB}{2}\!\right)\right]\nonumber\\&=&
\!\!\!\!\frac{i\pi}{s}\ .
\end{eqnarray} 
As discussed in Sec.~2, the solution is to derive an exceptional recursion
relation: the $B$-expansion relations. This is quite straightforward. We use
the $N=6$ recursion relation plus a correction term proportional
to $B$:
\begin{eqnarray}
\lefteqn{I(D;\nu_1,\nu_2,\nu_3,\nu_4)=}\nonumber \\
& &b_1 I(D;\nu_1-1,\nu_2,\nu_3,\nu_4)\nonumber \\
&+&b_2 I(D;\nu_1,\nu_2-1,\nu_3,\nu_4)\nonumber \\
&+&b_3 I(D;\nu_1,\nu_2,\nu_3-1,\nu_4)\nonumber \\
&+&b_4 I(D;\nu_1,\nu_2,\nu_3,\nu_4-1)\nonumber \\
&+&B(D+1-\sigma)I(D+2;\nu_1,\nu_2,\nu_3,\nu_4)
\end{eqnarray}
where 
\begin{equation}
I(D;\nu_1,\nu_2,\nu_3,\nu_4)=\int\frac{d^D l}{i\pi^{D/2}}
\frac{1}{d_1^{\nu_1}d_2^{\nu_2}d_3^{\nu_3}d_4^{\nu_4}}\ ,
\end{equation}
$\sigma=\sum\nu_i$ and $d_i=(l+q_i)^2-m_i^2$ the appropriate propagator.
This integral depends on the two Mandelstam invariants, $s$ and $t$.

This relation is exact for $N\leq 6$ in any momentum configuration. However,
unlike the usual recursion relations it has no termination point. For example
\begin{eqnarray}
\lefteqn{I(6;1,1,1,1)=}\nonumber\\&&
b_1 I(6-2\e;0,1,1,1)+b_2 I(6-2\e;1,0,1,1)\nonumber\\
&+&b_3 I(6-2\e;1,1,0,1)+b_4 I(6-2\e;1,1,1,0)\nonumber\\  
&+&(3-2\e)BI(8-2\e;1,1,1,1)\ .
\end{eqnarray}
The box integral $I(8-2\e;1,1,1,1)$ can be expressed in terms of 8-dimensional triangles
(proportional to $B$) and a 10-dimensional box integral (proportional to $B^2$).
By iterating this process, the scalar box is expressed as a 
series of triangles with increasing
factors of $B$. In the case of the massless 6-dimensional box we simply get:
\begin{eqnarray}
\lefteqn{I(6;1,1,1,1)}\nonumber \\ &=&
\frac{2}{s}\left[\sum_{m=0}^M B^m  a_mI(6+2m-2\e;0,1,1,1)\right] 
\nonumber\\  &+&
\frac{2}{t}\left[\sum_{m=0}^M  B^m a_mI(6+2m-2\e;1,0,1,1)\right] 
\nonumber\\  &+&
{\cal O}(B^{M+1})
\end{eqnarray}
with $a_m$ the expansion coefficients generated from the last term
in eq. 7.
In other words we can evaluate the 6-dimensional box integral in the forward
scattering region as an series expansion in $B$, with each coefficient 
given as  triangle integrals with one offshell leg. 
If the scattering angle is small,
not many iterations are needed for an accurate evaluation of 
the box diagram. This is illustrated in Fig.~1. Of course, for large $B$, the unmodified
integration by parts relations are completely stable.
\begin{figure}[t!]
\includegraphics[scale=0.75]{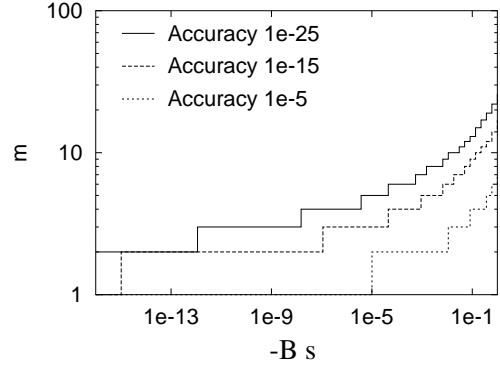}
\caption{The value of $-sB$ vs the depth of expansion $m$ 
needed to evaluate the scalar box integral $I(6;1,1,1,1)$ with a relative 
accuracy of $10^{-25}$, $10^{-15}$ and $10^{-5}$.}
\end{figure}

Note that the triangles appearing in the $B$-expansion relations
are all UV divergent. However, order by order in $B$ these divergences cancel in
the sums (because by construction the original integrals are UV finite.)
This means we can use a regulated expressions for $a_m$ and the triangles
integrals. A complication is that the factor $a_m$ itself is dependent
on the dimension. This means that the limit $\e\rightarrow 0$ has to be
taken after the UV divergences have been cancelled by adding different
triangles. An important property of the coefficient $a_m$ is that it 
is a function of the expansion depth $m$ and the value of $D-\sigma$
where both $D$ and $\sigma$ are given by the triangle the coefficient
is multiplying. 
The regulated expression to be used in the $B$-expansion relations is
given by
\begin{eqnarray}\lefteqn{J_m(D;\nu_1,\nu_2,\nu_3)\equiv
a_mI(D+2m;\nu_1,\nu_2,\nu_3)=}\nonumber\\&&\!\!\!\!\!\!
2^m\frac{(-1)^\sigma Q^{2m}}{\Gamma(\sigma+2n)}
\big(\nu_2\big)_n\big(\nu_3\big)_n\big(\sigma/2+n-m\big)_m
\nonumber\\ \!\!\!\!\!\!
&\times&\!\!\!\!\!\!\left(2\Psi(\sigma+2n)+\Psi(n+1)+\Psi(\sigma/2+n-m)
\right.\nonumber\\ &&\left.\!\!\!\!\!\!
-\Psi(\nu_2+n)-\Psi(\nu_3+n)-\Psi(\sigma/2+n)
\right.\nonumber\\ && \!\!\!\!\!\!\left.
-\log(-Q^2)+\gamma_E\right)
\end{eqnarray}
where $n$ is given by the relation $D=2(\sigma+n)$,
the Pochhammer's symbol $\big(n\big)_m=\Gamma(n+m)/\Gamma(n)$,
the Euler constant $\gamma_E$
and the Digamma function $\Psi(x)=d\,\log\big(\Gamma(x)\big)/d\,x$.
As indicated in the notation, in the algorithmic expansion we only
need to keep track of the tuple $(m,D,\nu_1,\nu_2,\nu_3)$ in order to
evaluate $J_m(D;\nu_1,\nu_2,\nu_3)$. For example, the $B$-expansion
of $I(8;2,1,1,1)$ using Eq.~(5) now becomes
\begin{eqnarray}
\lefteqn{I(8,2,1,1,1)=b_1\!\!\sum_{m=0}^M\!\! B^m\!J_m(8;0,1,1,1)}\nonumber\\
&+&\!\!\!\!
b_2\!\!\sum_{m=0}^M\!\!B^m\!\left[J_m(8;2,0,1,1)\!+\!b_1 J_m(8;1,0,1,1)\right]\nonumber\\
&+&\!\!\!\!
b_3\!\!\sum_{m=0}^M\!\!B^m\!\left[J_m(8;2,1,0,1)\!+\!b_1 J_m(8;1,1,0,1)\right]\nonumber\\
&+&\!\!\!\!  
b_4\!\sum_{m=0}^M\!\!B^m\!\left[J_m(8;2,1,1,0)\!+\!b_1 J_m(8;1,1,1,0)\right]\nonumber\\
&+&\!\!\!\! 
{\cal O}\left(B^{M+1}\right)\ .
\end{eqnarray}
Note that for $N=5$ the expansion becomes a bit more involved, but is still
rather straightforward.

\begin{figure}[t!]
\includegraphics[scale=0.75]{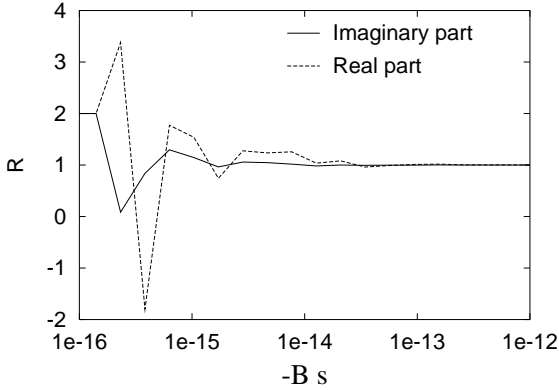}
\caption{The value of $-sB$ vs the ratio of the standard recursion result and the
$B$-expansion result for $I(6;1,1,1,1)$.}
\end{figure}
By looking at the ratio of the standard recursion relation and the 
$B$-expansion relation in Fig.~2 we can see clearly the numerical instability of
the standard recursion relation when $-sB<10^{-13}$. As can be seen from
Fig.~1
we already can neglect terms of order $B^3$ in the $B$-expansion
relation to achieve an accuracy of $10^{-25}$ in the integral evaluation.
Also shown, in Fig.~3 is the imaginary part of the scalar integral using both
recursion relations. As can be seen, the $B$-expansion relation can be
used over a large range of values provided the expansion depth is sufficient (as
indicited by Fig.~1).
\begin{figure}[t!]
\includegraphics[scale=0.75]{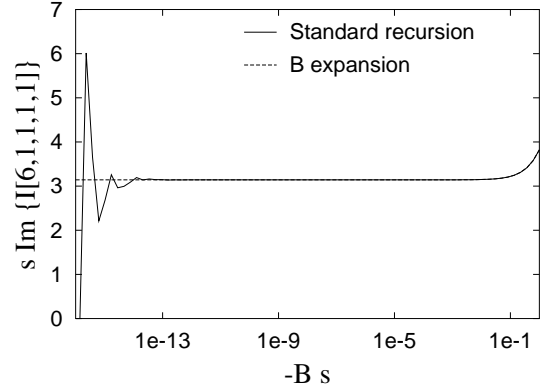}
\caption{The value of $-sB$ vs the imaginary part of the 6-dimensional box integral, $I(6,1,1,1,1)$.}
\end{figure}

We now can evaluate the $\gamma\gamma\rightarrow\gamma\gamma$ 
cross section in the forward region using both
the standard and $B$-expansion relations.
We define the normalized amplitude $A(\theta)$ 
\begin{equation}
{\cal M}(\theta)=\frac{e^4}{2\pi^2}A(\theta) 
\end{equation}
where $e$ is the coupling strength of the photon-fermion vertex.

In Fig.~4 we show the helicity amplitude for
$\gamma^+\gamma^+\rightarrow\gamma^-\gamma^-$ scattering.
In this case we have
\begin{equation}
A^{++;--}(\theta)=1\ .
\end{equation}
As can be seen in Fig.~4, the standard recursion relation has
a sudden loss in accuracy at around $\theta\approx 3\times 10^{-3}$.
On the other hand, the $B$ expansion formula gives the correct result
to within arbitrary precision.

Similarly, in Fig.~5 we show the helicity scattering  
$\gamma^+\gamma^+\rightarrow\gamma^+\gamma^+$.
The normalized helicity amplitude is now more complicated
\begin{eqnarray}
A^{++;++}(\theta)&=&
\frac{1+c^2}{4}\left[\log^2\left(\frac{1-c}{1+c}\right)+\pi^2\right]
\nonumber \\
&+& c\log\left(\frac{1-c}{1+c}\right)+1
\end{eqnarray}
where $c=\cos\theta$. The quantitative behaviour
is identical to the previous case; the $B$ improved recursion relations
give an accurate result in the domain where numerical instabilities render the
standard approach invalid.

In summary, the $B$-expansion method can be used to extrapolate
the algorithmic integration by parts method to all values of $\theta$
without any loss of accuracy. 
\begin{figure}[t!]
\includegraphics[scale=0.75]{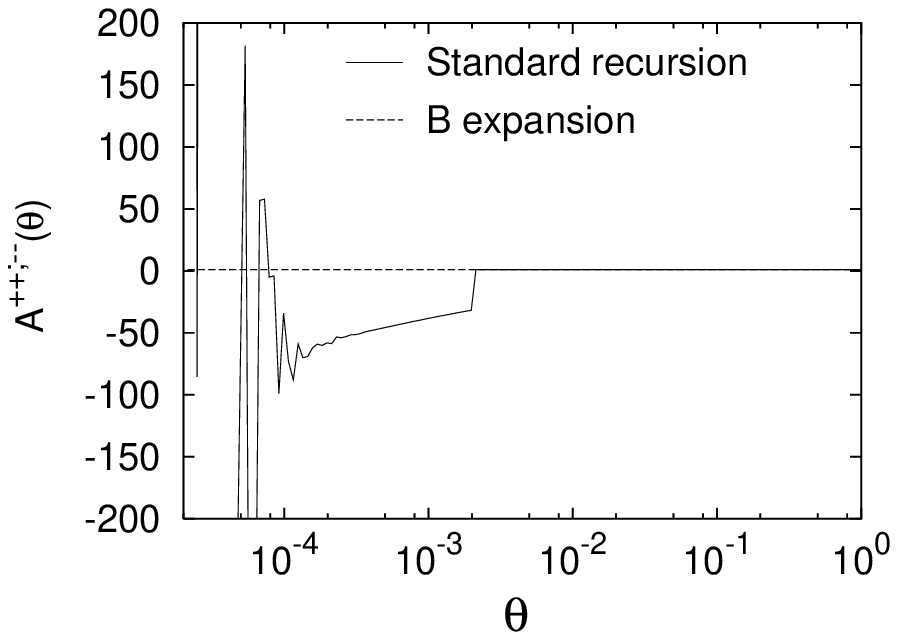}
\caption{The scattering angle vs the normalized amplitude for
$\gamma^+\gamma^+\rightarrow\gamma^-\gamma^-$.}
\end{figure}

\section{Conclusions} The method outlined can readily be generalized to
more complicated processes. As explained in sec. 2 this is done by
rewriting the recursion relations as expansion in the small $B$-parameter
or small eigenvalues of kinematic matrix. Because the recursion relations
are numerically solved for each phase space point, the method allows
exceptional kinematic configuration to be automatically detected by the
algorithm on an event-by-event basis. If needed the algorithm can decide
to use the expansion method without any knowledge  of the underlying
physics (e.g. threshold region, planar event configuration or other more
complicated configurations). This is highly desirable because of the
complexity of processes we are  ultimately interested in. For example,
four quark final states at hadron colliders such as $PP\rightarrow
t\bar{t}+b\bar{b}$ have multiple mass thresholds and numerous other
kinematic exceptional configurations.  Similarly processes like
$P\overline P\rightarrow W+4$ jets requiring the evaluation of the complicated
one-loop diagrams involving six partons plus a vector boson for which the
exceptional kinematic configurations are difficult to comprehend.

The algorithm  outlined in these proceedings
augments the integration by parts algorithm of Ref.~\cite{GiGlo}. The combined
numerical procedure should be able to calculate arbitrarily complicated
one-loop amplitudes in {\it all} regions of phase space with arbitrary precision.

\begin{figure}[t!]
\includegraphics[scale=0.75]{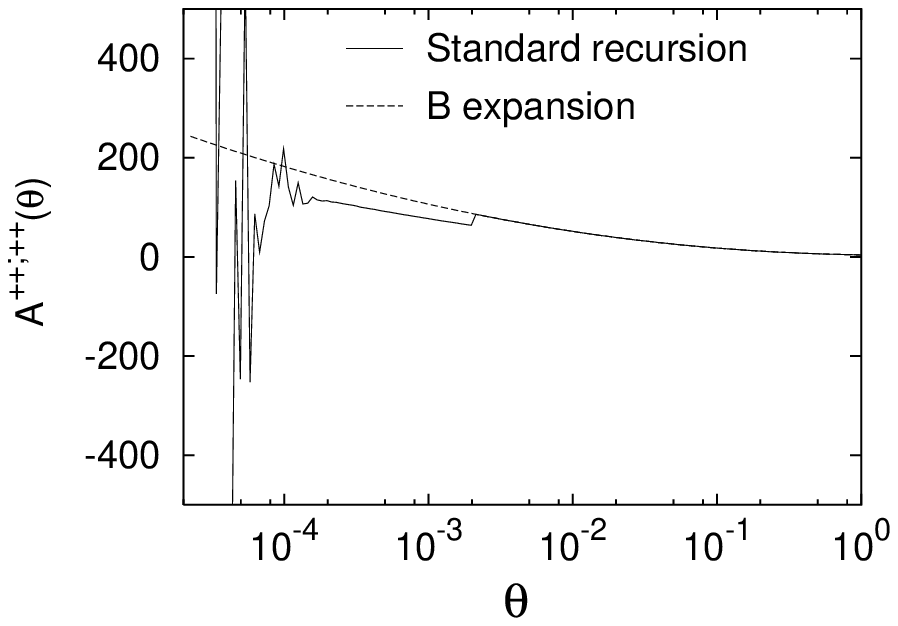}
\caption{The scattering angle vs the normalized amplitude for
$\gamma^+\gamma^+\rightarrow\gamma^+\gamma^+$.}
\end{figure}


\begin{thebibliography}{9}
\bibitem{CheTka} 
K.~G.~Chetyrkin and F.~V.~Tkachov, Nucl. Phys. B 192 (1980) 159;
Z.~Bern, L.~J.~Dixon and D.~A.~Kosower, Nucl. Phys. B 412 (1994) 751;
J.~Fleicher, F.~Jegerlehner and O.~V.~Tarasov, Nucl. Phys. B 566 (2000) 423;
T.~Binoth, J.~P.~Guillet and G.~Heinrich, Nucl. Phys. B 572 (2000) 362;
G.~Duplancic and B.~Nizic, Eur. Phys. J. C35 (2004) 105;
F.~del Aguila and R.~Pittau, arXiv:hep-ph/0404120.
\bibitem{GiGlo} W.T. Giele and E.W.N. Glover, JHEP 0404, 029 (2004).
\bibitem{CaGloMi} J.M. Campbell, E.W.N. Glover and D.J. Miller, Nucl. Phys. B 498 (1997) 397.
\bibitem{FePaParUc} A. Ferroglia, M. Passera, G. Passarino and S. Uccirati, Nucl. Phys. B 650 (2003) 162.
\end{thebibliography}
\end{document}